\documentclass[aps,pra,showpacs]{revtex4}

\usepackage{graphicx}

\newcommand{\ie}{\emph{i.e.}}
\newcommand{\eg}{\emph{e.g.}}

\newcommand{\Er}{\ensuremath{E_\mathrm{r}}}
\newcommand{\wrecoil}{\ensuremath{\omega_\mathrm{r}}}
\newcommand{\precoil}{\ensuremath{p_\mathrm{r}}}

\newcommand{\scold}{\ensuremath{\sigma_{\mathrm{cold}}}}
\newcommand{\shot}{\ensuremath{\sigma_{\mathrm{hot}}}}
\newcommand{\Tcold}{\ensuremath{T_\mathrm{cold}}}

\newcommand{\e}{\ensuremath{\mathrm{e}}}

\newcommand{\un}[1]{\, \mathrm{#1}}

\begin{document}

\title{Time dependence of laser cooling in optical lattices}

\author{Claude M. Dion}
\email{claude.dion@tp.umu.se}

\author{Peder Sj\"{o}lund} 
\author{Stefan J. H. Petra}
\author{Svante Jonsell}
\author{Anders Kastberg} 

\affiliation{Department of Physics, Ume{\aa} University, SE-901\,87 Ume{\aa},
  Sweden}

\begin{abstract}
  We study the dynamics of the cooling of a gas of caesium atoms in an
  optical lattice, both experimentally and with 1D full-quantum Monte
  Carlo simulations.  We find that, contrary to the standard
  interpretation of the Sisyphus model, the cooling process does not
  work by a continuous decrease of the average kinetic energy of the
  atoms in the lattice.  Instead, we show that the momentum of the
  atoms follows a bimodal distribution, the atoms being gradually
  transferred from a hot to a cold mode.  We suggest that the cooling
  mechanism should be depicted in terms of a rate model, describing
  the transfer between the two modes along with the processes
  occurring within each mode.
\end{abstract}


\maketitle

\section{Introduction}

Laser cooling is now a well-established technique enabling to reach
temperatures of the order of the microkelvin in a gas of
atoms~\cite{cool:nobel98}.  It has made possible Bose-Einstein
condensation~\cite{bec:nobel02}, while laser-cooled atoms are used,
for instance, in atomic clocks~\cite{cold:bize05} and as frequency
standards~\cite{cold:hollberg05}, and have been proposed for parity
violation measurements~\cite{cold:sanguinetti03} and quantum
information processing~\cite{qi:monroe02}.

An appropriate superposition of laser beams can result in a
spatially-periodic modulation of the polarisation of the light,
creating an optical
lattice~\cite{lat:jessen96,lat:guidoni99,lat:grynberg01}.  A
multilevel atom with a degenerate ground state moving through such a
lattice will experience large non-adiabatic effects as the light shift
of the different substates and the optical pumping rates then depend
on the position of the atom.  The standard model used to explain laser
cooling in optical lattices is Sisyphus
cooling~\cite{cool:castin91,cool:cohen-tannoudji92}, where an atom
will preferentially jump from the crest of a potential hill to the
valley of an other potential, losing kinetic energy each time.  Atom
dynamics can be expressed in a semi-classical way by this model as the
interplay between a velocity-dependent cooling force due to the
Sisyphus effect and a diffusion term describing the recoil from photon
absorption-emission cycles along with fluctuations in the gradient
force when going from one potential curve to the
other~\cite{cool:castin91,cool:cohen-tannoudji92}.

Although it was predicted many years ago~\cite{cool:castin91}, it is
only recently that experimental observations have shown non-Gaussian
velocity distributions in optical lattices~\cite{lat:jersblad04}, as
well as in atomic fountains~\cite{cool:sortais00}.  While the standard
Sisyphus model considers a cooling force linear in
velocity~\cite{cool:cohen-tannoudji92}, leading to a Gaussian momentum
distribution, refinements using non-linear
forces~\cite{cool:castin91,cool:hodapp95} give non-Gaussian
distributions, following for instance Tsallis or L\'{e}vy
statistics~\cite{lat:lutz03,cool:bardou02}.  Results obtained by
taking into account the localisation of the atoms around the potential
minima also show deviations from a
Gaussian~\cite{cool:javanainen92,cool:molmer93}.  Moreover, arguments
based on a ``band'' picture of optical lattices~\cite{lat:castin91},
where some atoms are trapped in the potential wells of the lattice
while others move around in the ``conduction band'', call for a
bimodal distribution~\cite{lat:sanchez-palencia02,lat:jersblad04}.
Indeed, it was found that a double Gaussian function provides the best
fit to both experimental and simulated data for the steady state
momentum distribution of atoms~\cite{lat:jersblad04}.

We report in this letter a study of the time evolution of the cooling
of a gas of caesium atoms inside an optical lattice.  It provides new
elements to understand the origin of the non-Gaussian momentum
distributions.  Apart from a few studies concerning
localisation~\cite{lat:raithel97a,lat:raithel97b},
diffusion~\cite{lat:carminati01}, and damping
rates~\cite{lat:carminati03,lat:sanchez-palencia03b} of atoms in
optical lattices, this is one of the first observation of the dynamics
of the cooling process.

\section{Methodology}


Details of the experimental setup can be found in
refs.~\cite{lat:jersblad00,lat:ellmann01}. Briefly, we first
accumulate $^{133}$Cs atoms in a magneto-optical trap.  We adjust the
irradiance and the detuning, then we turn off the magnetic field and
leave the atoms in an optical molasses with even further reduced
irradiance and increased detuning.  The atoms are thus cooled to about
$40\un{\mu K}$ before being transfered to the optical lattice; such a
``hot'' initial cloud of atoms is used to reduce the effects from the
initial kick given to the atoms when the lattice is turned on.  The
lattice itself has a 3D structure made up from a four-beam
configuration~\cite{lat:jessen96,lat:grynberg01}: two laser beams are
linearly polarised along the $x$ axis and propagate in the $yz$ plane
symmetrically with respect to the $z$ axis, whereas the other two
beams are polarised along the $y$ axis and propagate in the $xz$ plane
symmetrically with respect to $z$. This yields a tetragonal pattern of
points with pure circular polarisation, alternately $\sigma^+$ and and
$\sigma^-$. The lasers are slightly detuned below the resonance of the
$F_{\mathrm{g}} =4 \rightarrow F_{\mathrm{e}} =5$ transition of the D2
line ($6\mathrm{s}^2\mathrm{S}_{1/2} \rightarrow
6\mathrm{p}^2\mathrm{P}_{3/2}$) at $\lambda = 852\un{nm}$.  After a
given time the optical lattice is turned off (in $4.2\un{\mu s}$) and the
atoms are left to fall and then detected by a probe beam located $\sim
5\un{cm}$ below the trapping region.  This turn-off time is fast
enough so that no adiabatic cooling of the atomic cloud can occur. The
momentum distribution of the atoms is recovered from the
time-of-flight signal.


The simulation of the dynamics of the atoms in the optical lattice is
carried out using a full-quantum Monte Carlo wave-function
method~\cite{mc:castin95} for the $F_{\mathrm{g}} =4 \rightarrow
F_{\mathrm{e}} =5$ transition of caesium~\cite{lat:jersblad04}.  Due to
the numerical complexity of the 3D case, we restrict ourselves to a 1D
lin~$\perp$~lin configuration, reproducing the same alternation of
$\sigma^+$ and $\sigma^-$ potential wells (see, \eg,
ref.~\cite{lat:guidoni99}).  The time dependence of the cooling
process is obtained by considering a series of different histories
where the initial momentum of the atom is chosen randomly from a
normal distribution corresponding to a temperature of $50\un{\mu K}$.
Observables are recovered by averaging over these histories, \eg, the
momentum distribution is obtained from the time-dependent wave
functions $\psi_h$ by
\begin{equation}
D(p;t) = \frac{1}{N} \sum_{h=1}^{N} \left| \psi_h(p;t) \right|^2 \,,
\label{eq:dist}
\end{equation}
where the index $h$ labels the different histories.

\section{Results}

\begin{figure}[t]
\centerline{\includegraphics[width=0.5\textwidth]{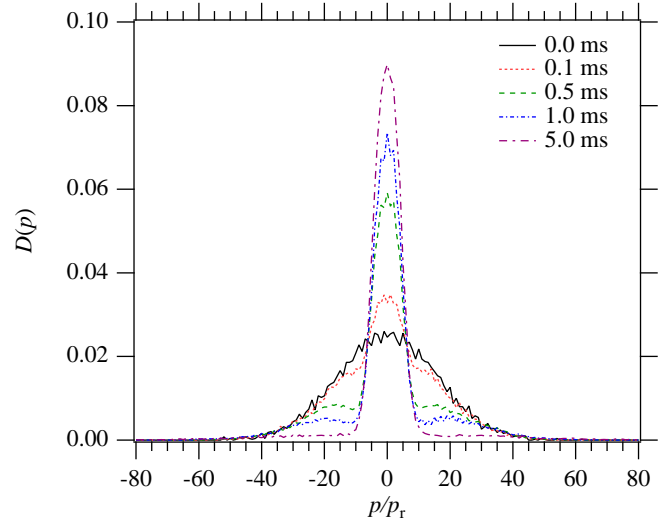}}
\caption{Simulated momentum distribution [see eq.~(\ref{eq:dist})] at
  different lattice times, for $\Delta = -10 \Gamma$ and $U_0 = 127
  \Er$.}
\label{fig:td_simul}
\end{figure}
\begin{figure}[b]
\centerline{\includegraphics[width=0.48\textwidth]{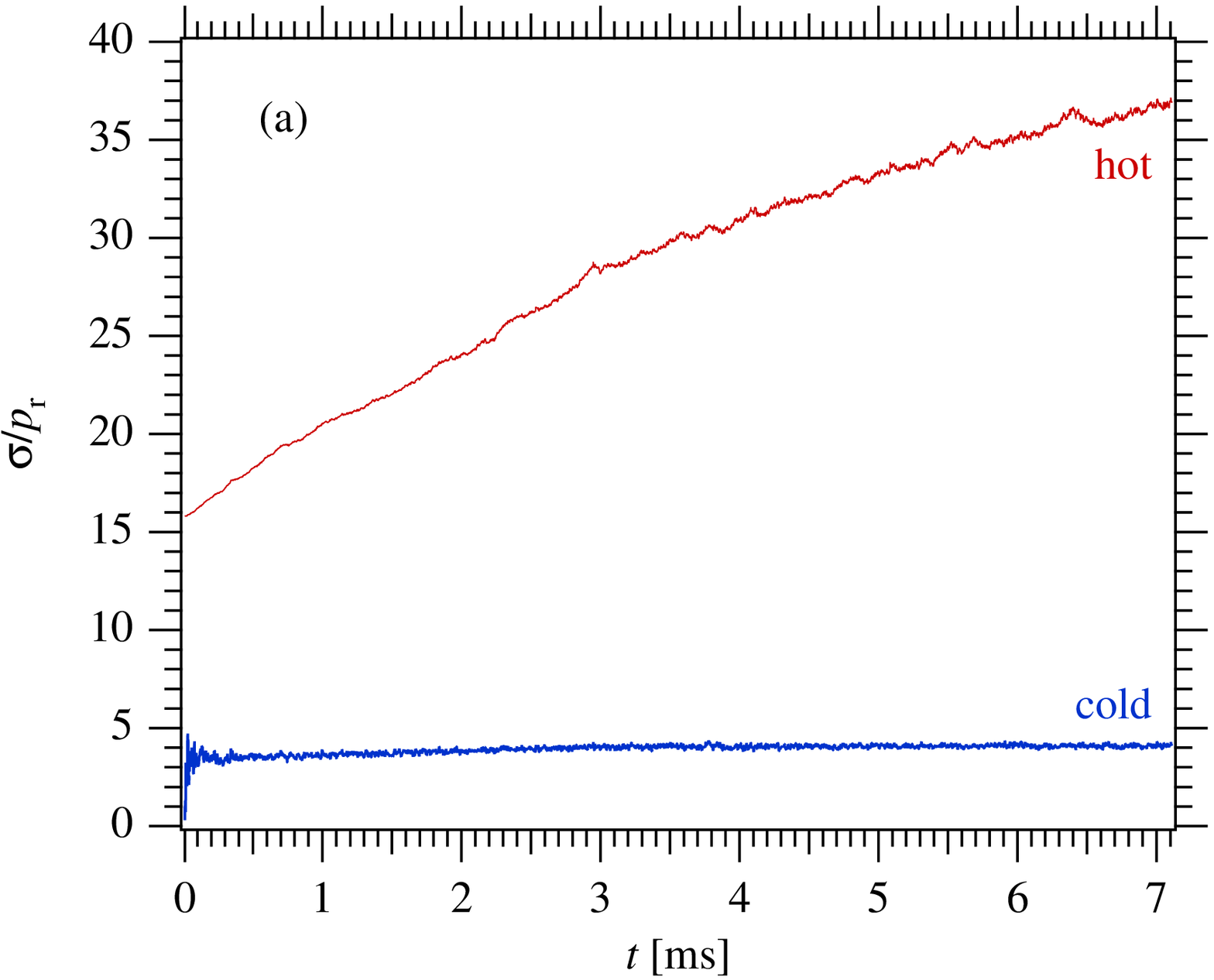} 
\includegraphics[width=0.48\textwidth]{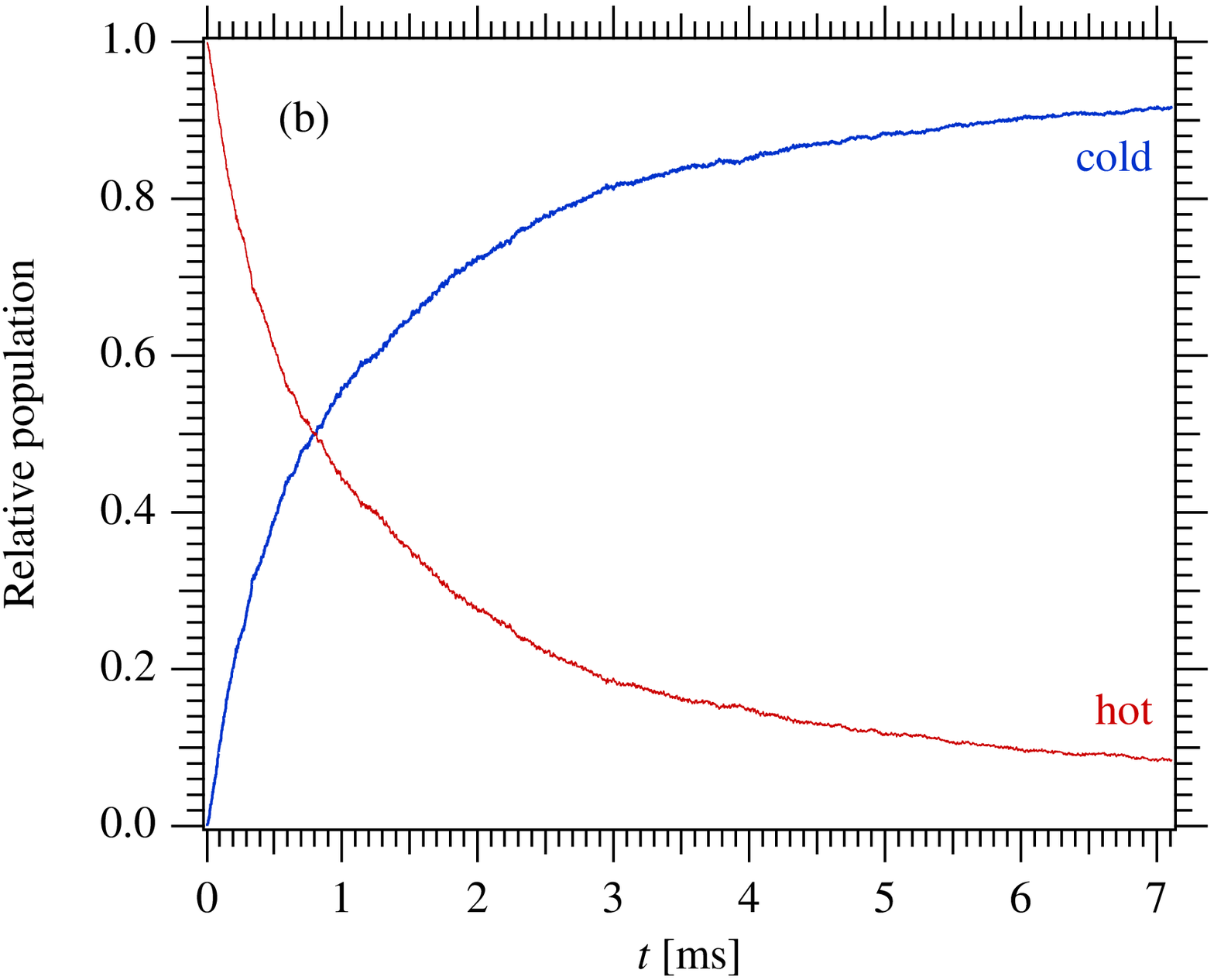}}
\caption{Time dependence of the Gaussian fit parameters [see
  eq.~(\ref{eq:dg})] for the cold and hot modes, for the simulated
  data shown in fig.~\ref{fig:td_simul}: (a)~widths $\scold$ and
  $\shot$; (b)~relative populations.}
\label{fig:td_fit}
\end{figure}

Sample momentum distributions obtained by the numerical simulation are
given in fig.~\ref{fig:td_simul}, for a detuning of $\Delta = -10
\Gamma$ and a potential well depth of $U_0 = 127 \Er$, with $\Gamma/2
\pi = 5.2227\un{MHz}$ the natural linewidth of the
$6\mathrm{p}^2\mathrm{P}_{3/2}$ level of caesium and $\Er = \hbar
\wrecoil$ the recoil energy, where $\wrecoil/2\pi = 2.0663\un{kHz}$ is
the atomic recoil frequency~\cite{alkali:steck03}.
The momentum is expressed in units of the recoil momentum $\precoil =
h/\lambda$. Starting from the initial distribution at $50\un{\mu K}$,
we clearly see that a narrow peak gradually grows in the centre of the
momentum distribution.  At the same time, the population of the wings
from the initial distribution decreases and spreads out to higher
momentum.  It appears that the atoms are not gradually cooled but
populate progressively the central feature: they are transfered from a
``hot'' to a ``cold'' mode.  To analyse these results, we have fitted
the momentum distribution to a double Gaussian,
\begin{equation}
  D(p) = A_{\mathrm{cold}} \exp\left[ -p^2 / \left(2
      \scold^2 \right) \right] + A_{\mathrm{hot}}
  \exp\left[ -p^2 / \left(2 \shot^2 \right) \right]\,,
  \label{eq:dg}
\end{equation}
by finding the parameters ($A$ and $\sigma$) that best reproduce the
second, $\left\langle p^2 \right\rangle$, and fourth, $\left\langle
  p^4 \right\rangle$, moments of the distribution while conserving the
norm, with the constraint $\scold < \shot$.  This method appears
better suited than the traditional $\chi^2$ fitting procedure to
reproduce the wings of the distribution when the cold mode dominates.
We find, fig.~\ref{fig:td_fit}(a),
that there is no time evolution of the width $\scold$, hence we can
assign a temperature $\Tcold = \scold^2 / (m k_{\mathrm{B}})$ to the
cold mode, where $m$ is the mass of the atom and $k_{\mathrm{B}}$ the
Boltzmann constant. This is another striking feature of the cooling
process, which works not only by transferring atoms between the two
modes, but puts them in an \emph{unchanging} cold mode.  On the other
hand, atoms in the hot mode are continuously heated.  We have checked
that no steady state is obtained in the hot mode for longer times.
For as long as we can see, the momentum distribution in the hot mode
shows a Gaussian profile in the region where it can clearly be
distinguished from the cold mode.  In contrast, as can be seen in
fig.~\ref{fig:td_fit}(b), the relative population of each mode
(calculated as $\sqrt{2 \pi} A \sigma$ for a normalised distribution)
reaches a steady state after some time.  By fitting the population of
the cold mode to an exponential function, we recover a transfer rate
of $\tau^{-1} \approx 7.9\times 10^{2}\un{s^{-1}}$.  This rate is
about 15 times slower than the localisation rate previously measured
for caesium in a 1D lattice~\cite{lat:raithel97a} and 2 orders of
magnitude slower than the scattering rate.  The persistence of hot
atoms seems to be due to the fact that some higher momentum atoms are
heated and never cooled. Indeed, starting from a lower temperature
such as $5\un{\mu K}$ leads to a greater transfer from the hot to the
cold mode, with less heating in the hot mode.

\begin{figure}
\centerline{\includegraphics[width=0.48\textwidth]{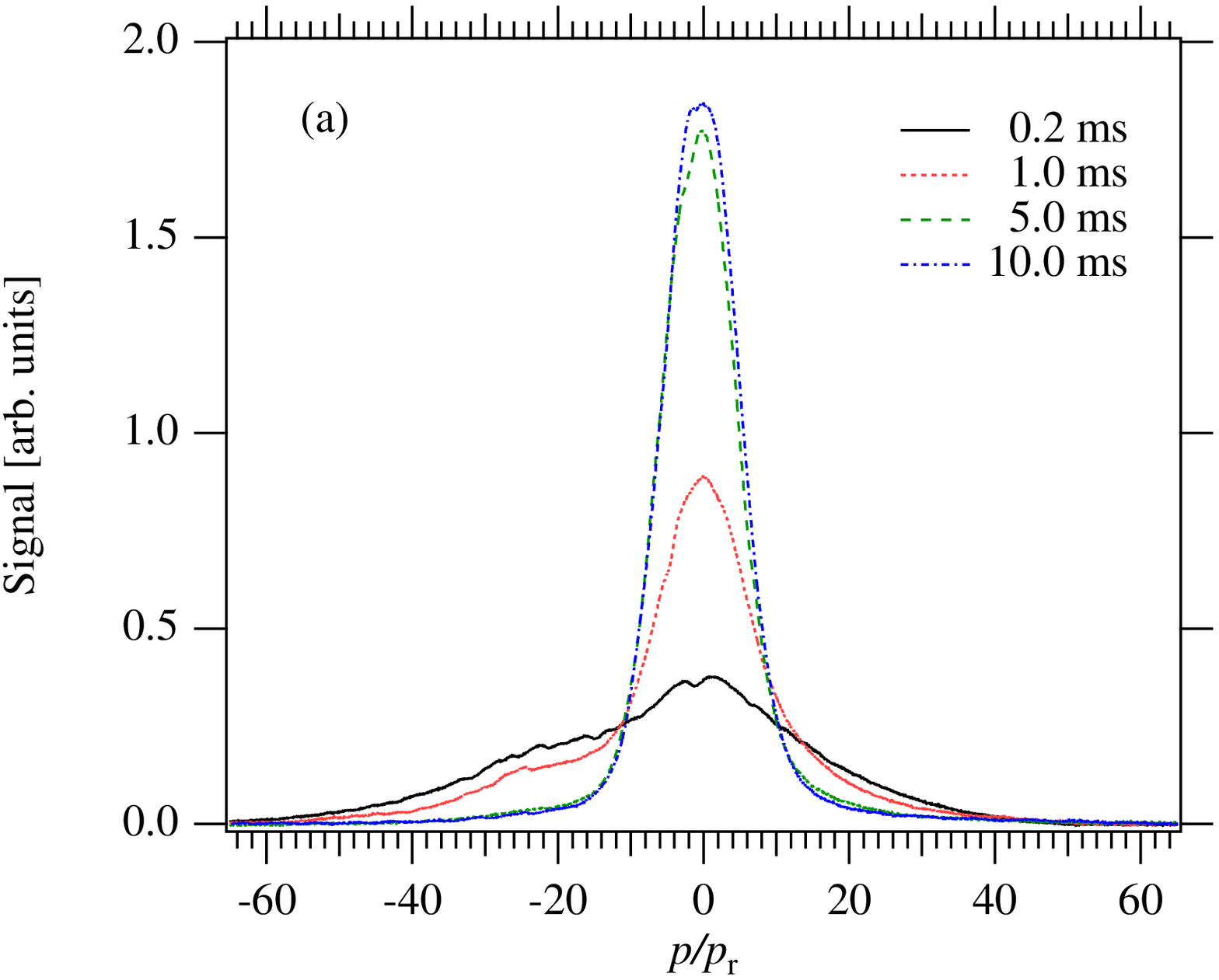}
\includegraphics[width=0.48\textwidth]{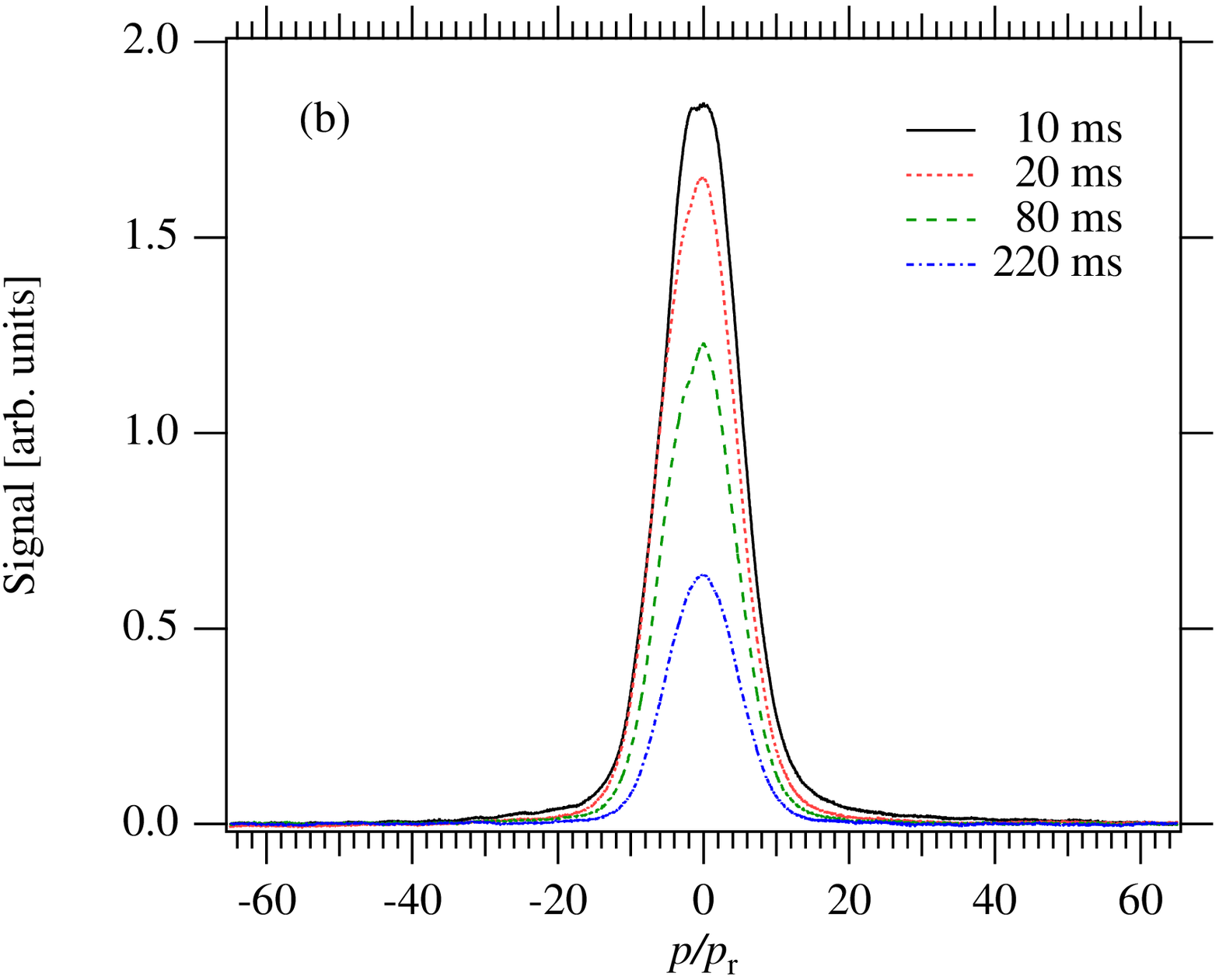}}
\caption{Experimental momentum distribution at different lattice
  times, for $\Delta = -12.6 \Gamma$ and $U_0 = 217 \Er$.}
\label{fig:td_exp}
\end{figure}

Turning now to the experimental results, we show in
fig.~\ref{fig:td_exp}
the momentum distribution obtained for a laser detuning of $\Delta =
-12.6 \Gamma$ and an irradiance corresponding to a potential depth
$U_0 = 217 \Er$. (Since the simulation is for a 1D lattice, a
quantitative comparison to the experiment is not meaningful, so no
attempt is made to exactly match the parameters for both cases.)  We
see again the appearance of a central feature of almost constant
width.  The main difference with the simulation is that after a
certain time the signal decreases, as seen in
fig.~\ref{fig:td_exp}(b).  This is due to an overall loss of atoms due
to their diffusion out of the lattice and collisions with background
gas, while the numerical simulation considers an effectively infinite
lattice.  This phenomenon is probably also responsible for another
difference between the two results, namely the absence in the
experimental signal of a significant heating of the hot mode, as the
high momentum atoms seen in the simulation would quickly escape from
the lattice region.

\begin{figure}
\centerline{\includegraphics[width=0.48\textwidth]{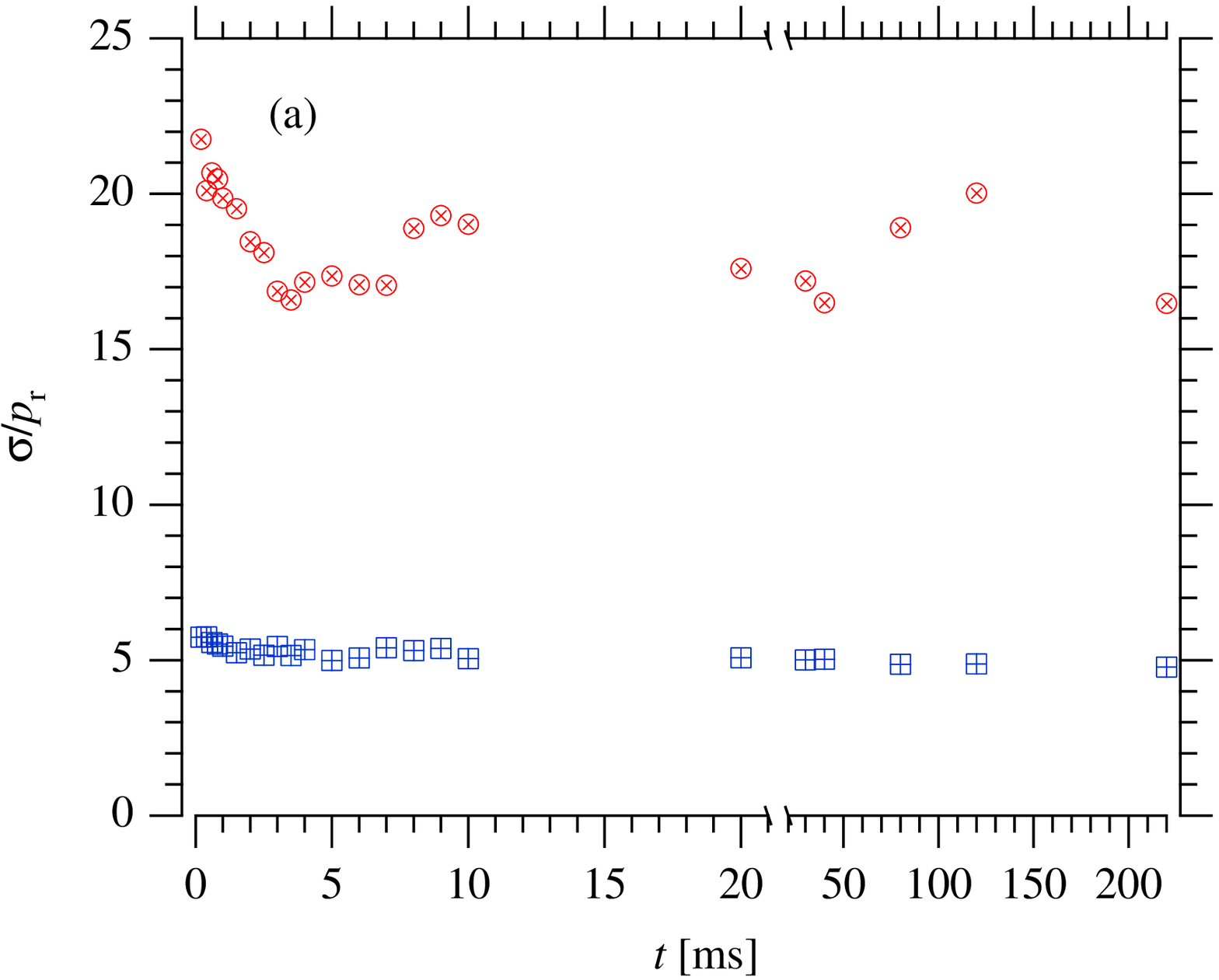}
\includegraphics[width=0.48\textwidth]{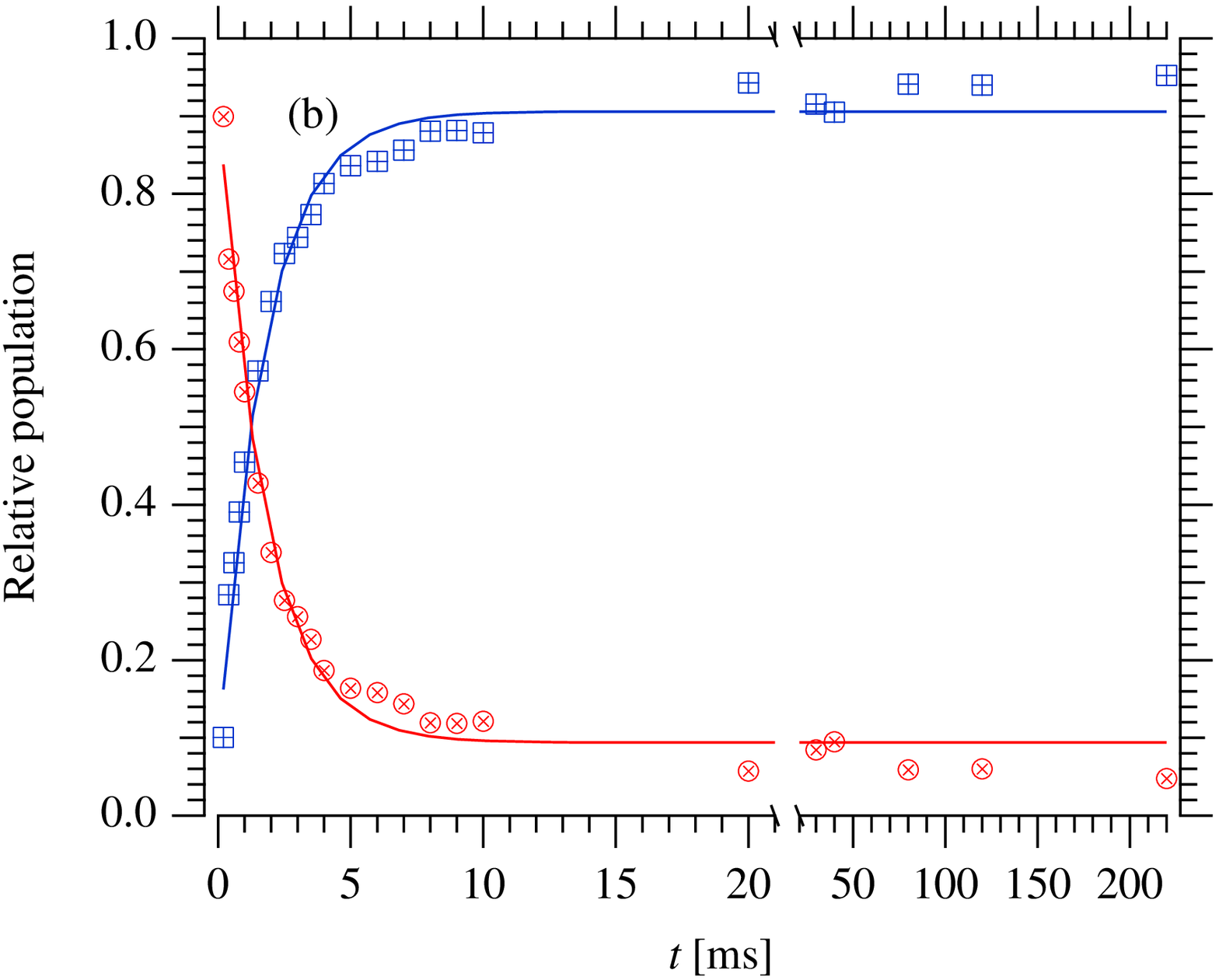}}
\caption{Time dependence of the Gaussian fit parameters [see
  eq.~(\ref{eq:dg})] for the cold (squares) and hot (circles) modes,
  for the experimental data shown in fig.~\ref{fig:td_exp}. (a)~Widths
  $\scold$ and $\shot$. (b)~Relative populations, with fits to
  exponential functions (solid lines).  Note that there is a scale
  change in the time axis.}
\label{fig:td_fit_exp}
\end{figure}

The result obtained by fitting the experimental data to a double
Gaussian is given in fig.~\ref{fig:td_fit_exp}.
Where have reverted here to a $\chi^2$ method as a fit using the
moments of the distribution gave aberrant results due to the bump
present in the experimental data at negative momentum.  It is clear
from the time evolution of $\shot$ (although it is noisy) that there
is no heating observed in the hot mode.  This is nevertheless not
necessarily contradictory with the result of the simulation, as the
decrease $\shot$ could be explained by the preferential loss of the
high momentum atoms.  The temperature in the cold mode is almost
constant, the slight decrease with time being within the experimental
and fitting errors.  The relative populations of the two modes shown
in fig.~\ref{fig:td_fit_exp}(b) seem to indicate that a steady state
is not reached, with the cold mode continuously gaining in population,
which would indicate that the loss of atoms is through the hot mode.
The rate of transfer to the cold mode is found to be of the same order
as for the simulation, with $\tau^{-1} \approx 5.8 \times
10^{2}\un{s^{-1}}$, which is 6 times slower than the localisation rate
in a 3D lattice~\cite{lat:raithel97a} and $\sim 400$ times slower than
the scattering rate.

\begin{figure}
\centerline{\includegraphics[width=0.48\textwidth]{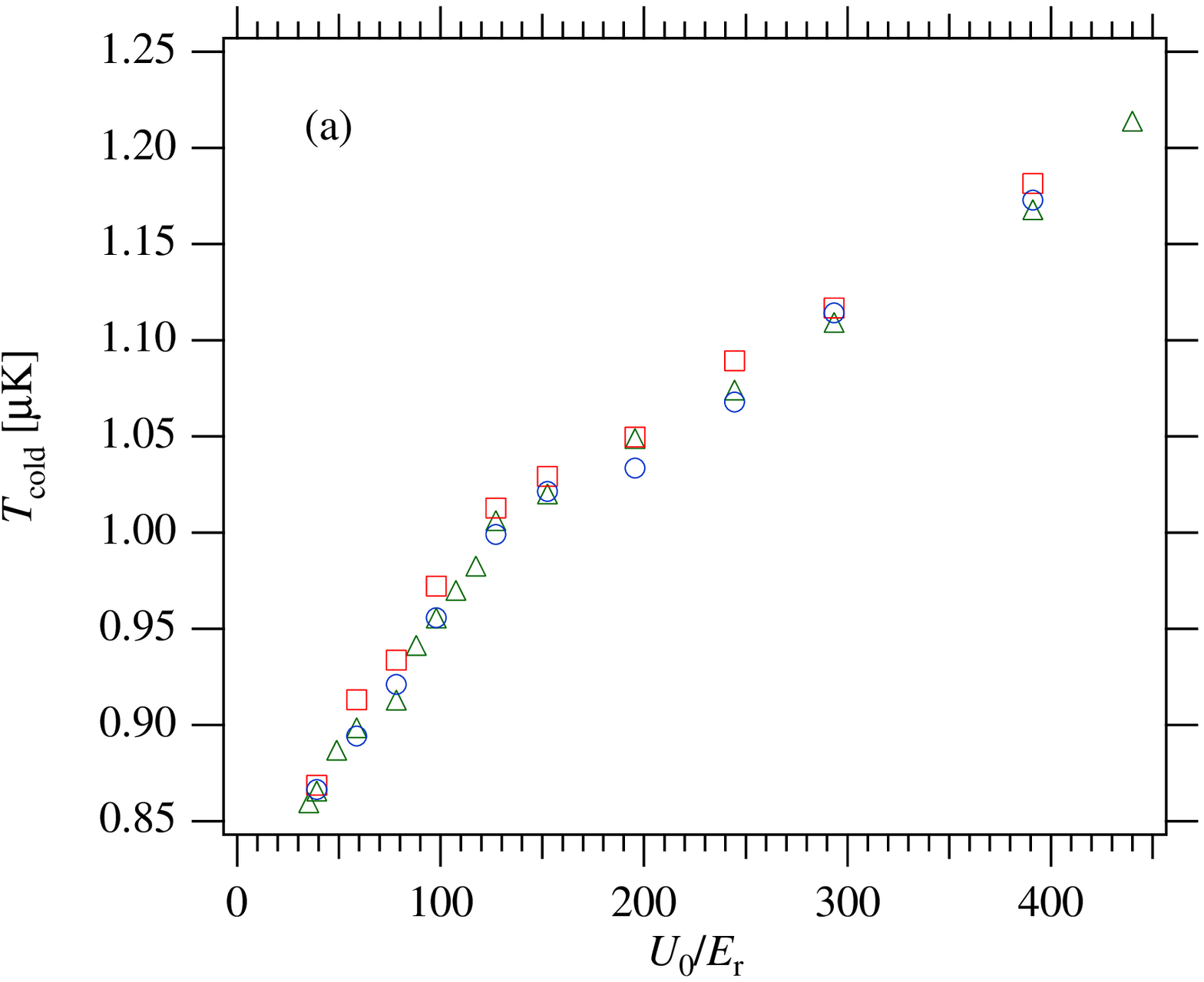}
\includegraphics[width=0.48\textwidth]{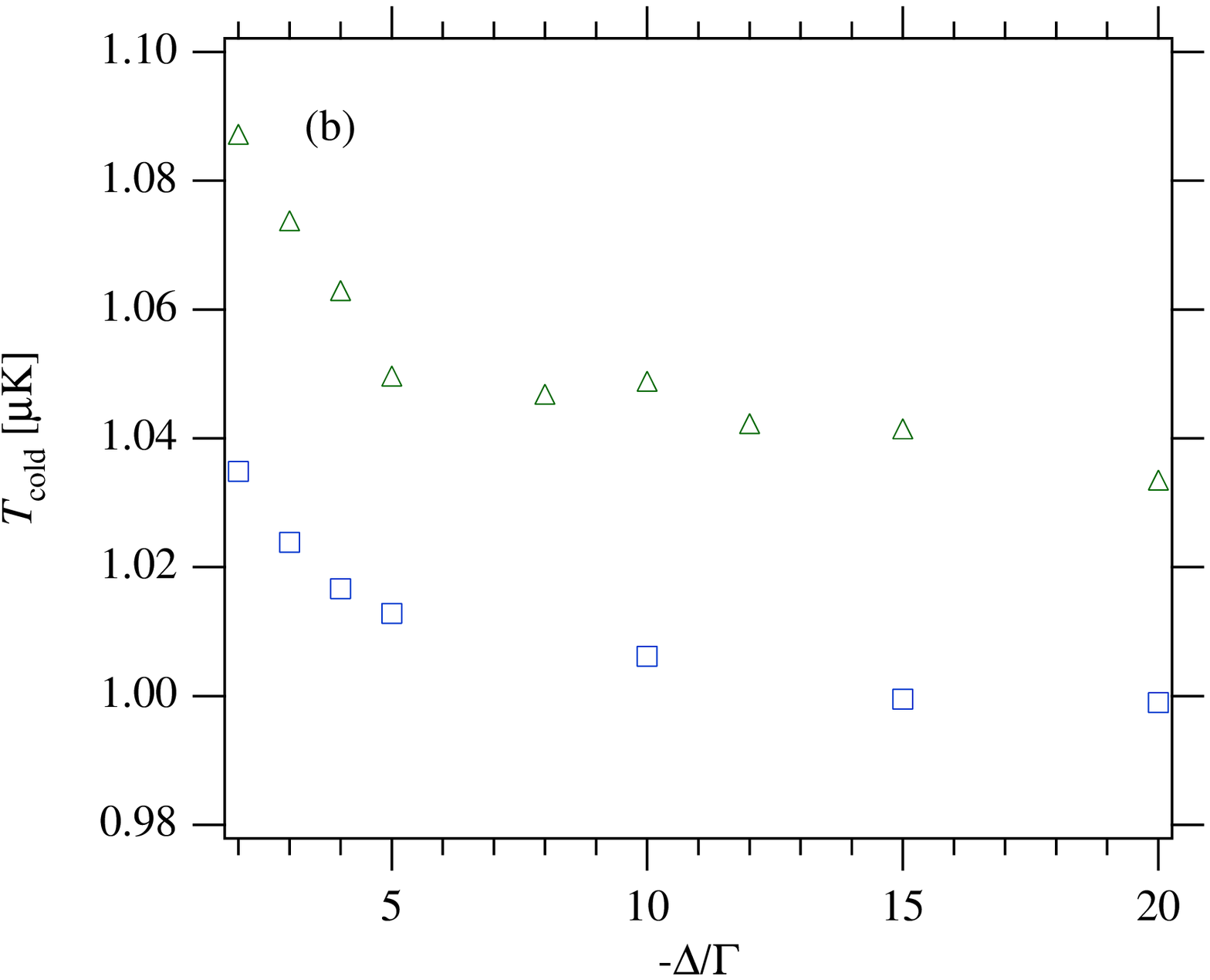}}
\caption{Temperature of the cold mode: (a) as a function of the
  potential depth, for detunings of $-5 \Gamma$ (squares), $-10
  \Gamma$ (triangles), and $-20 \Gamma$ (circles); (b) as a function
  of detuning, for well depths of $127\Er$ (squares) and $196\Er$
  (triangles).}
\label{fig:temperature}
\end{figure}

We ran simulations over a range of the parameters $\Delta$ and $U_0$
and the temperatures thus obtained for the cold mode are given in
fig.~\ref{fig:temperature}. 
The variation of $\Tcold$ as a function of the well depth looks
similar to that obtained in previous
studies~\cite{lat:castin91,lat:carminati01,lat:jersblad04}, although
the temperature was then calculated from the root-mean-square
momentum.  It also appears that the temperature does not follow the
linear dependence with potential depth predicted by an analytical
model of Sisyphus cooling~\cite{cool:dalibard89} for the entire range
considered here.  We also note that no \emph{d\'{e}crochage} is seen, \ie,
there is no threshold potential below which there would be a rapid
increase of the temperature.  The curve of
fig.~\ref{fig:temperature}(a) is in fact similar to the results
previously obtained by considering not the r.m.s.\ momentum, but the
width of the distribution at
$1/\sqrt{\e}$~\cite{lat:castin91,lat:jersblad04}.  The dominance of
the cold mode over the hot mode is such that the width at
$1/\sqrt{\e}$ is essentially equal to $\scold$, except for very short
times.  It thus seems that the previously observed d\'{e}crochage appears
when the population of the hot mode is great enough to influence
significantly the r.m.s. momentum.  It is a feature of the relative
population of the modes, not of the cooling temperature reached.  
Extrapolating to a vanishing potential depth results in a non-zero
value of $\Tcold$, but we expect the behaviour to become non linear as
the number of bound states in the lattice goes to zero.
The
temperature of the cold mode also varies with the detuning, as shown
in fig.~\ref{fig:temperature}(b).  This is consistent with what has
been observed for
rubidium~\cite{lat:carminati01,lat:sanchez-palencia03b}, although it
appears that $\Tcold$ decreases exponentially with $| \Delta |$,
instead of following the functional form proposed in
refs.~\cite{lat:carminati01,lat:sanchez-palencia03b}.

\section{Discussion}

Our results call for a discussion of the cooling of atoms in optical
lattices in terms of rates for a bimodal distribution, based on the
hypothesis that the cold and hot modes correspond to trapped and
untrapped atoms, respectively. The dominant process during the initial
cooling stage is the transfer from the hot to the cold mode.  Within
the hot mode, simulations show that there is an important heating
effect, while for the cold mode basically no heating or cooling is
seen, which means that there is a fast equilibrium reached inside the
cold mode.  This latter process depends on the detuning of the lasers,
as seen by the dependence of $\Tcold$ not only on the potential depth
but also on the detuning.  There is also the transfer from the cold to
the hot mode, which, according to simulations, is slow with respect to
the transfer from hot to cold.  Otherwise, there would be a gradual
depletion of the cold mode as atoms in the hot mode are heated to the
extent where they cannot go back to the cold mode.  Nevertheless,
under the hypothesis that cold atoms are trapped, the loss of atoms
due to diffusion could only be explained by a coupling from cold to
hot.  We also note that associating the cold mode with trapped atoms
is not incompatible with the observation that localisation is faster
than cooling, as atoms can spend more time near the potential minima
(\ie, localise) while still being untrapped.  In addition, the rate
actually measured in ref.~\cite{lat:raithel97a} may also reflect the
progressive filling of lattice sites.

\section{Conclusion}

To summarise, we have shown that atoms in an optical lattice follow a
bimodal distribution of hot and cold atoms.  The cooling inside the
lattice takes place via a transfer of the atoms from the hot to the
cold mode.  We suggest that the cooling of atoms in an optical lattice
be depicted in terms of a rate model, describing the transfer between
the two modes along with the processes occurring within each mode.
This calls for further investigations in order to understand by
exactly which mechanism this is taking place.

\acknowledgments 

We thank R. Kaiser, E. Lutz, K. M{\o}lmer, and L. Sanchez-Palencia for
stimulating discussions. This work was supported by Carl Tryggers
stiftelse, Magnus Bergvalls stiftelse, Kempestiftelserna, K \& A
Wallenbergs stiftelse, and the Swedish Research Council.
S.J.H.P. thanks Carl Tryggers stiftelse for financial support.  This
research was conducted using the resources of the High Performance
Computing Center North (HPC2N).



\clearpage


\end{document}